\documentclass[useAMS,usenatbib]{mn2e}

\usepackage{graphicx}                   %for PS/EPS graphics inclusion, new
\usepackage{float}
\usepackage{amsmath}
\usepackage{amssymb}
\usepackage{multirow}
\usepackage{bm}
\usepackage{color}

\input{epsf.sty}                        %for PS/EPS graphics inclusion, old
\input{psfig.sty}
\begin{document}

\title[$H$-cluster stars]
{$H$-cluster stars}

\author[Lai, Gao \& Xu]{X. Y. Lai,$^{1,2}$ C. Y. Gao$^2$ and R. X. Xu$^2$\\
$^1$School of Physics, Xinjiang University, Urumqi 830046, China\\
$^2$School of Physics and State Key Laboratory of Nuclear Physics
and Technology, Peking University, Beijing 100871, China}

\maketitle

\begin{abstract}
The study of dense matter at ultra-high density has a very long
history, which is meaningful for us to understand not only cosmic
events in extreme circumstances but also fundamental laws of
physics.
It is well known that the state of cold matter at supra-nuclear
density depends on the non-perturbative nature of quantum
chromo-dynamics (QCD) and is essential for modeling pulsars.
A so-called $H$-cluster matter is proposed in this paper as the
nature of dense matter in reality.

In compact stars at only a few nuclear densities but low
temperature, quarks could be interacting strongly with each other
there.
That might render quarks grouped in clusters, although the
hypothetical quark-clusters in cold dense matter has not been
confirmed due to the lack of both theoretical and experimental
evidence.
Motivated by recent lattice QCD simulations of the $H$-dibaryons
(with structure $uuddss$), we are therefore  considering here a
possible kind of quark-clusters, $H$-clusters, that could emerge
inside compact stars during their initial cooling, as the dominant
components inside (the degree of freedom could then be $H$-clusters
there).
Taking into account the in-medium stiffening effect, we find that at
baryon densities of compact stars $H$-cluster matter could be more
stable than nuclear matter.
 We also find that for the $H$-cluster matter with lattice structure,
the equation of state could be so stiff that it would seem to be
``superluminal'' in most dense region.
However, the real sound speed for $H$-cluster matter is in fact hard to
calculate, so at this stage we do not put constraints on our model
from the usual requirment of causality.

We study the stars composed of $H$-clusters, i.e., $H$-cluster
stars, and derive the dependence of their maximum mass on the
in-medium stiffening effect, showing that the maximum mass could be
well above 2 $M_\odot$ as observed and that the resultant
mass-radius relation fits the measurement of the rapid burster under
reasonable parameters.
Besides a general understanding of different manifestations of
compact stars, we expect further observational and experimental
tests for the $H$-cluster stars in the future.

\end {abstract}

\section{Introduction}
\label{Introduction}

Possible matter at the highest density, {\em limited by the sizes of
electrons and nuclei}, was incidentally speculated in a seminal
paper~\citep{fowler1926} about a decade after Rutherford suggested
his model of the atom.
Compact objects, especially at density as high as nuclear matter
density, are gradually focused on by astronomers and physicists, the
study of which opens a unique window that relates fundamental
particle physics and astrophysics.
It is worth noting that above the saturated nuclear matter density, $\rho_0$, the
state of matter is still far from certainty, whereas it is essential
for us to explore the nature of pulsars.
Historically, at average density higher than $\sim 2\rho_0$, the
quark degrees of freedom inside would not be negligible, and such
compact stars are then called quark
stars~\citep{Itoh:1970uw,Haensel1986,Alcock:1986hz}.
Bodmer-Witten conjecture says that strange quark matter (composed of
$u$, $d$ and $s$ quarks) could be more stable than
nuclear matter~\citep{Bodmer:1971we, Witten:1984rs}.
Although the effect of non-perturbative QCD makes it difficult to
derive the real state of cold quark matter, the existence of quark
stars cannot be ruled out yet, neither theoretically nor
observationally~\citep[see a review in][]{Weber2005}.

Although quark matter at high density but low temperature is
difficult to be  created in laboratory as well as difficult to be
studied by pure QCD calculations, some efforts have been made to
understand the state of cold quark matter and quark stars.
MIT bag model treats the quarks as relativistic and weakly
interacting particles, which is the most widely used model for quark
stars~\citep{Haensel1986,Alcock:1986hz}.
The color super-conductivity (CSC) state is currently focused on
perturbative QCD as well as QCD-based effective
models~\citep{Alford2008}.
In most of these models, quark stars are usually characterized by
soft equations of state, because the asymptotic freedom of QCD tells
us that as energy scale goes higher, the interaction between quarks
becomes weaker and weaker.

In cold quark matter at realistic baryon densities of compact stars
($\rho\sim 2-10\rho_0$), however, the energy scale is far from the
region where the asymptotic freedom approximation could apply.
In this case, the interaction energy between quarks could be
comparable to the Fermi energy, so that the ground state of
realistic quark matter might not be that of Fermi gas~\citep[see a
discussion given in][]{Xu:2010}.
Some evidence in heavy ion collision experiments also shows that the
interaction between quarks is still very strong even in the case of hot
quark-gluon plasma~\citep{Shuryak2009}.
It is then reasonable to infer that quarks could be coupled strongly
also in the interior of those speculated quark stars, which could
make quarks to condensate in position space to form quark
clusters.~\footnote{Besides this top-down scenario, i.e., an
approach from deconfined quark state with the inclusions of stronger
and stronger interaction between quarks, one could also start from
hadronic state (a bottom-up scenario): strangeness may play an
important role in {\em gigantic} nuclei so that the degree of
freedom would not be nucleon but quark-cluster with strangeness.}
The observational tests from polarization, pulsar timing and
asteroseismology have been discussed~\citep{Xu03}, and it is found
that the idea of clustering quark matter could provide us a way to
understand different manifestations of pulsars.
The realistic quark stars could then be actually ``quark-cluster
stars''.~\footnote{Strictly speaking, quark-cluster stars are {\em
not} quark stars if one thinks that the latter are composed by free
quarks. Nevertheless, in this paper, we temporarily consider
quark-cluster stars as a very special kind of quark stars since (1)
both kinds of stars are self-bound and quark-cluster stars manifest
themselves similar to quark stars rather than gravity-bound neutron
stars, and (2) the quark degree of freedom plays a significant role
in determining the equation of state and during the formation of
quark-cluster stars.}
An interesting suggestion is that quark matter could be in a solid
state~\citep{Xu03,Horvath05,Owen05,mrs07}, and for quark-cluster
stars, solidification could be a natural result if the kinetic
energy of quark clusters is much lower than the interaction energy
between the clusters.

Quark clusters may be analogized to hadrons, and in fact some
authors  have discussed the stability of hadron bound states.
A dihyperon with quantum numbers of $\Lambda\Lambda$ ($H$ dibaryon)
was predicted to be a stable state or resonance~\citep{Jaffe1977},
and an 18-quark cluster (quark-alpha, $Q_{\alpha}$) being completely
symmetric in spin, color and flavor spaces was also
proposed~\citep{Michel1988}.
$H$ dibaryon in lattice QCD simulations, although no direct evidence
from experiments, provides us a specific kind of quark clusters that
could be very likely to exist inside quark stars.
In fact, $H$ dibaryons have been studied for years as a possible
kind of multi-quark compound states.
The non-relativistic quark-cluster model was introduced to study
the binding energy of $H$-clusters~\citep{Straub:1988mz}.
The interaction between $H$-clusters was investigated by employing
one-gluon-exchange potential and an effective meson exchange
potential, and a short-range repulsion was
found~\citep{Sakai:1997jf}.
Recently, $H$ dibaryon, with binding energy of about 10 to 40
MeV, has been found in lattice QCD simulations by two independent
groups~\citep{Beane:2010hg,Inoue:2010es}, and STAR preliminary
results show also possible stable $H$ dibaryons from
$\Lambda$-$\Lambda$ correlations (Huanzhong Huang, private
communications).

Strange quark matter, with light flavor symmetry, has nearly
equal numbers of $u$, $d$ and $s$ quarks.
If quark-clusters are the dominant components of strange quark
matter, then it is natural to conjecture that each quark-cluster
could composed of almost equal numbers of $u$, $d$ and $s$ quarks.
During the initial cooling of a quark star,
the interaction between quarks will become stronger and stronger,
then $H$-clusters (six-quark clusters with the same structure as
$H$ dibaryons $uuddss$) would emerge from the combination of
three-quark clusters (with structure $uds$ as $\Lambda$ particles),
due to the attraction between them.
If the light flavor symmetry is ensured, then the dominant
components inside the stars is very likely to be $H$-clusters.
In our previous work about the quark-clusters stars, the number of
quarks inside each quark-cluster $N_q$ is taken to be a free
parameter~\citep{LX09b}, and as the further study in this paper, we
realistically specify the quark-clusters to be $H$-clusters.

There could be other kind of particles with strangeness, such as
kaons and hyperons.
Kaon condensate would probably reduce the maximum mass of the stars
and hyperons heavier than $\Lambda^0$ would not have large enough
number densities, both of which would not have significant effect on
the stars' global structure.
We neglect them in this paper as the first step towards the
structure of this specific kind of quark-cluster stars, and the
effect of all possible particles should be taken into account in our
further studies.

To study the $H$-cluster stars, we assume that the interaction
between $H$-clusters is mediated simply by $\sigma$ and $\omega$
mesons and introduce the Yukawa potential to describe the $H$-$H$
interaction~\citep{Faessler1997}, and then derive the mass-radius
relations of $H$-cluster stars in different cases of the in-medium
stiffening effect and surface density.
Under a wide range of parameter-space, the maximum mass of
$H$-cluster  stars can be well above 2$M_\odot$, and therefore such
compact stars cannot be ruled out even though some pulsars with mass
as high as 2$M_\odot$ are found~\citep{Lai:2010wf}.
Moreover, the observations (e.g. pulsar-mass) could help us
constrain the $H$-$H$ interaction in dense matter.

The paper is arranged as follows. The existence and localization of
$H$-clusters inside compact stars are discussed in \S  2. The
equation of state and the global structure of $H$-cluster stars are
given in \S 3, including the dependence of their maximum mass on
the in-medium stiffening effect and surface density of the star.
Some issues about the $H$-cluster stars are discussed in \S 4, and
we make conclusions in \S 5.

\section{$H$-clusters inside compact stars}

The state of matter of compact stars is essentially a problem of
non-perturbative QCD, with energy scale below 0.8 GeV (corresponding
to mass density below $10\rho_0$).
If the interaction between quarks could be strong enough to group
them into clusters, the quark-clustering phase should be very
different from the CSC phase under perturbative QCD, and it could
also be different from the normal hadron phase if the quark matter has
light flavor symmetry.
Whether $H$-clusters could be the dominant component inside compact
stars is an interesting but difficult problem, and here we just make
some rough estimation about their existence and quantum effect.
We find that at densities inside pulsars, the in-medium effect could
make $H$-cluster matter to be more stable than nucleon matter, where
$H$-clusters are stable against decaying to nucleons.
Moreover, they could be localized rather than that of Bose-Einstein
condensation, and such a localization of quark-clusters could lead
to a classical crystalline structure to form solid state.
In addition, the lattice vibration would not dissolve $H$-clusters.

\subsection{The stability of $H$-cluster matter}

Whether the Bodmer-Witten conjecture is true or not is hard for us to
solve from the first principle.
Here we demonstrate that $H$-cluster matter could be stable with respect
to transforming into nucleon matter at the same density, by assuming
the Brown-Rho scaling.

In dense matter (ignoring the mass-difference between neutrons and
protons), the masses of neutrons
and mesons satisfy the scaling law
$m_N^*/m_N=m_M^*/m_M$, where $m_N$
and $m_M$ denote the mass of nucleons and mesons, and the masses
with and without asterisks stand for in-medium values and free-space
values respectively.
This is called Brown-Rho scaling~\citep[for a review, see][]{BR2004}.
We suppose that the Brown-Rho scaling holds for $H$-dibaryons, which
have the same mass-scaling as nucleons
\begin{equation}
m_N^*/m_N=m_M^*/m_M=m_H^*/m_H=1-\alpha_{BR}\frac{n}{n_0}, \label{BRh}
\end{equation}
where $n$ denotes the number density of $H$-dibaryons,
$n_0$ denotes the number density of saturated nuclear matter,
and $\alpha_{BR}$ is the coefficient of scaling.
For nuclear matter, $\alpha_{BR}$ is found to be 0.15 for mesons and 0.2
for nucleons by fitting experimental data~\citep{brown1991}.
The density dependence of $\alpha_{BR}$ is still unknown, especially at
supra-nuclear density.
In our following calculations, we treat $\alpha_{BR}$ as a parameter in the
range between 0.1 and 0.2, and for simplicity we assume that its value is
the same for both mesons and $H$-dibaryons.

At first, for simplicity we do not consider the interaction of nucleons when
compare the energy per baryon of nuclear matter and $H$-cluster matter.
For the system composed of nucleons in weak equilibrium at densities
higher than $\rho_0$, the dominant component is neutrons.
Taking into account the in-medium effect, the energy per baryon of
neutron matter is $E/A=\sqrt{p_n^2+m_n^{*2}}+E_{sym}$,
where $n_n$ is number density of neutrons, $p_n=(3\pi^2 n_n)^{1/3}$
is the Fermi momentum of
neutrons, $E_{sym}$ is the symmetry energy per baryon, and we use the
expression $E_{sym}=31.6(n_n/n_0)^{1.05}$ MeV~\citep{chen2005}.
We also consider the nucleon matter with equal number density of neutrons,
protons and electrons, $n_n=n_p=n_e$, where the symmetry energy
is vanishing but the electron Fermi energy $p_e=(3\pi^2 n_e)^{1/3}$
is high, and the energy per baryon is
$E/A=\sqrt{p_n^2+m_n^{*2}}+p_e/2$.
The energy per baryon of $H$-cluster matter is $E/A=\epsilon/n_B$, where
$\epsilon$ is the energy density of $H$-cluster matter calculated in
\S 3.1, and $n_B$ is the baryon number density ($n_B=2n$).
We set free-space value for the mass of $H$-cluster
to be $m_H=2m_\Lambda-20\ \rm MeV=2210\ \rm MeV$.

\begin{figure}
\includegraphics[width=3 in]{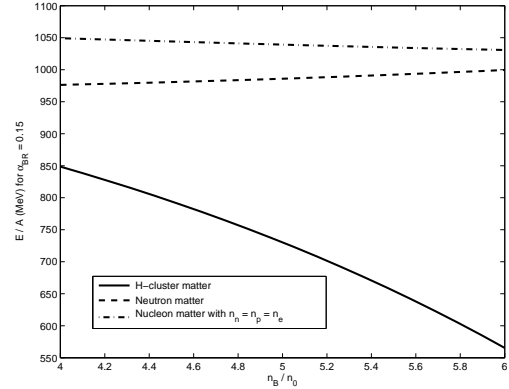}
\caption{The dependence of energy per baryon $E/A$ on
baryon number density $n_B$, for
$H$-matter (solid line), neutron matter (dashed line) and nucleon
matter with $n_n=n_p=n_e$ (dash-dotted line), in the case
$\alpha_{BR}=0.15$. It is evident that $H$-particles would be stable
when the number density is above $2n_0$.
\label{figEA}}
\end{figure}

Figure~\ref{figEA} shows the dependence of $E/A$ on baryon number density $n_B$,
in the cases $\alpha_{BR}=0.15$, for $H$-matter, neutron matter
and nucleon matter with
$n_n=n_p=n_e$.
We can see that $H$ matter is more stable than nuclear
matter when the number density is larger than $2n_0$ (due to the in-medium
effect, the rest-mass energy density of $H$-cluster matter at $2n_0$ is about
$3.3\rho_0$).
If the stability condition is satisfied, $H$-clusters would not
decay to nucleons, where the in-medium effect play the crucial role
in stablizing $H$-cluster matter.
The decreasing of the mass of $H$-dibaryons with increasing
densities could be equivalent to the increasing of binding energy of
$H$-dibaryons, which makes $H$-clusters to be more stable.

%%%%%%%%%%%%%%%%%%%%%%%%%%%%%%%%%%%%%%%%
%%%%%%%%%%%%%%%%%%%%%%%%%%%%%%%%%%%%%%%%

\begin{figure}
\includegraphics[width=3 in]{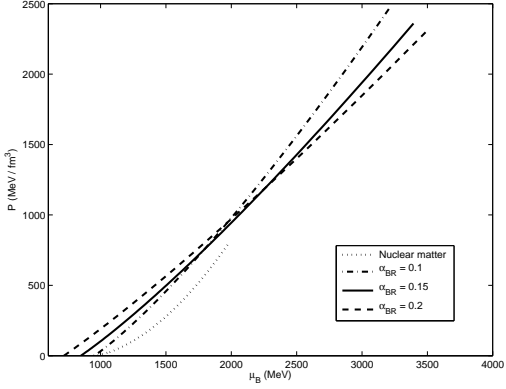}
\caption{The chemical potential for baryon $\mu_B$, for nuclear
matter~\citep{NG2010} (dotted line), and $H$-cluster matter with
$\alpha_{BR}=0.1$ (dash-dotted line),
$\alpha_{BR}=0.15$ (solid line) and $\alpha_{BR}=0.2$ (dashed line),
where the surface density $n_s=2n_0$.
For $H$-cluster matter the densities are
chosen in the range where the stars could be gravitationally stable,
and for nuclear matter the densities are chosen in the range
as the widest range of baryon number densities among the three
(with $n_B\simeq 8.3 n_0$).
\label{figMu}}
\end{figure}

Actually, the stability condition we discussed above is applied to
the surface of $H$-cluster stars, where the pressure is vanishing,
and the densities we choose above can be seem as the surface
densities (given the surface density we can get coefficients in the
interaction potential, see \S 3.1).
At high densities, the interaction between nucleons would
become significant to resist gravity.
To compare the stability of $H$-cluster matter and
nuclear matter at high densities inside stars,  we choose one of
the models which describe the state of nuclear matter~\citep{NG2010}.
We compare the chemical potential per baryon
$\mu_B=(\epsilon+P)/n_B$ of $H$-cluster matter with that of nuclear
matter~\citep{NG2010}, when pressure is non-zero.
The results are shown in Figure~\ref{figMu}, where for $H$-cluster
matter, in three cases $\alpha_{BR}=0.1$, 0.15 and 0.2, where the
surface density $n_s=2n_0$, and the densities are
chosen in the range where the stars could be gravitationally stable
(according to Figure~\ref{figMR}), and for nuclear matter the
densities are chosen in the range as the widest range among the three,
with $n_B\simeq 8.3 n_0$.
We can see that $H$-cluster matter
would be more energetically favorable than nuclear matter when
$\mu_B\lesssim2000$ MeV.
Whether there would be phase transition from $H$-cluster matter to
some other forms of matter at higher densities is unknown because
of the ignorance of state of matter at densities beyond several times
of $n_0$, so we do not extrapolate the curve of nuclear matter
to larger $\mu_B$ in Figure~\ref{figMu} and consider the possible phase
transition from this figure.
As discussed in \S 4.5, although the state of dense matter at ultra-high
density is uncertain, 3-flavor symmetry may result in a ground state of
matter.

%%%%%%%%%%%%%%%%%%%%%%%%%%%%%%%%%%%%%%%%
%%%%%%%%%%%%%%%%%%%%%%%%%%%%%%%%%%%%%%%%

%
%For nuclear matter we choose the model

\subsection{Crystallization of $H$-cluster matter}

Under the interaction, $H$-clusters could be localized and behave
like classical particles.
In the core of a neutron star, $H$-clusters could also appear, and
the existence of $H$-clusters inside neutron stars has been studied
in relativistic mean-field theory~\citep{Faessler1997}. It was found
that when the potential between $H$-clusters is negative enough,
then a substantial number density of $H$-clusters will reduces the
maximum mass of neutron stars if Bose-Einstein condensation
happens~\citep{Glendenning:1998wp}.

\begin{figure}
\includegraphics[width=3 in]{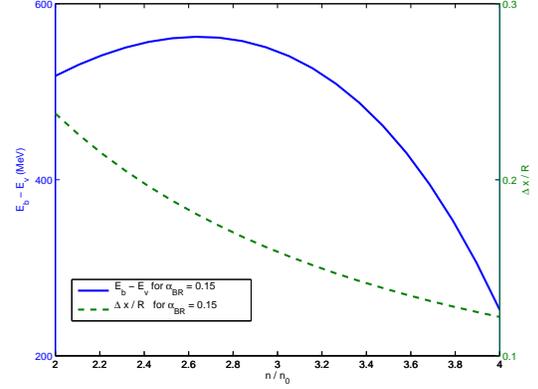}
\caption{The comparison of binding energy $E_b$ and vibration energy
$E_v$ (solid line), and the ratio of the deviation from the
equilibrium position $\Delta x$ to the distance between two nearby
$H$-clusters $R$ (dashed line), as the function of number density $n$,
in the case $\alpha_{BR}=0.15$. If $E_b-E_v>0$, $H$-clusters in
crystalline structure would be stable against lattice vibration; if
$\Delta x / R<1$, the quantum effect would not be significant and
the Bose-Einstein condensate would not take place.
\label{figEb}}
\end{figure}

However, due to the strong interaction, $H$-clusters would be
localized like classical particles in crystal lattice, and the
quantum effect would be negligible.
One $H$-cluster is under the composition of interaction from its
neighbor $H$-clusters, which forms a potential well.
The energy fluctuation makes this $H$-cluster oscillate about its
equilibrium position with the deviation $\Delta x$, i.e., the
vibration energy $E_v$, can be derived as $E_v \simeq\hbar ^2/(m_H
\Delta x^2)\simeq k \Delta x^2$, where $k\simeq
\partial^2V(r)/\partial r^2$, and $r$ is the distance of two
neighbor $H$-clusters.
We use the $H$-$H$ interaction in Eq(\ref{V}), and estimate $E_v$
and $\Delta x$ at different densities.
Taking $\alpha_{BR}=0.15$ as an example, the results are shown in
Figure~\ref{figEb}.
The density range we choose here is narrow, since we will see in \S
3.2 that the density of a stable star will be well below 4
$\rho_0$ in the case $\alpha_{BR}=0.15$.
In the proper density range the vibration energy is smaller than the
binding energy of $H$-clusters, which means that the fluctuations
about the lattice would not dissolve $H$-clusters.
On the other hand, the distance between two nearby $H$-clusters
$R=n^{-1/3}$ (with $n$ the number density of $H$-clusters) is larger
than $\Delta x$, which means that the quantum effect could not be
significant and the Bose-Einstein condensate would not take place.

%%%%%%%%%%%%%%%%%%%%%%%%%%%%%%%%%%%%%%%%
%%%%%%%%%%%%%%%%%%%%%%%%%%%%%%%%%%%%%%%%

$H$-clusters are localized because each of them feels an
ultra-strong repulsion every direction around it, and such
localization could lead to a crystalline structure. In fact, the
relation between hard-core potential and crystallization was
discussed previously~\citep[e.g.,][]{Canuto:1975kr}. The almost
infinitely strong repulsion is certainly an ideal case, but in real
world the short range of $H$-$H$ interaction could be still strong
enough to localize them.

\section{The global structure of $H$-cluster stars}

We propose a possible kind of quark-cluster stars totally composed
of $H$-clusters, i.e., $H$-cluster stars. $H$-cluster stars could
have different properties from neutron stars and conventional quark
stars, such as the radiation properties, cooling behavior and global
structure. In this paper, we only focus on the global structure of
$H$-cluster stars, deriving the mass-radius relation based on the
equation of state.

\subsection{$H$-$H$ interaction and equation of state}

The interaction between $H$-clusters has been studied under the
Yukawa potential with $\sigma$ and $\omega$
coupling~\citep{Faessler1997}, and we adopt this form of interaction
here
\begin{equation}
V(r)=\frac{g_{\omega H}^2}{4\pi}\frac{e^{-m_\omega
r}}{r}-\frac{g_{\sigma H}^2}{4\pi}\frac{e^{-m_\sigma
r}}{r},\label{V}
\end{equation}
where $g_{\omega H}$ and $g_{\sigma H}$ are the coupling constants
of $H$-clusters and meson fields.
The numerical result of the potential between two $H$-dibaryons
shows a minimum at  $r_0\approx 0.7$ fm with the depth $V_0\approx
-400$ MeV~\citep{Sakai:1997jf}, which means that, to get the minimal
point, two $H$-dibaryons should be very close to each other.

Nevertheless, the medium effect in dense matter could change those
properties.
In dense nuclear matter, the effective meson masses $m_M^*$ satisfy
the Brown-Rho scaling law of Eq.(\ref{BRh})~\citep{BR2004}
This shows the in-medium effect that stiffens the inter-particle
potential by reducing the meson effective masses, and $m_\sigma$ and
$m_\omega$ in Eq.(\ref{V}) should be replaced by $m_\sigma^*$ and
$m_\omega^*$.
In addition, the mass of $H$-dibaryons $m_H$ also obeys the same
scaling law.

Given the potential between two $H$-clusters, we can get the energy
density by taking into account all of the contributions from
$H$-clusters in the system.
In the case of a strong repulsive core, each $H$-cluster could be
trapped inside the potential well as demonstrated before.
Assuming the localized $H$-clusters form the simple-cubic structure,
from Eq.(\ref{V}) we can get the interaction energy density
$\epsilon_{I}$ as the function of the distance between two nearby
$H$-clusters $R$
\begin{equation}
\epsilon_{I}=\frac{1}{2}n\left(A_1\frac{g_{\omega
H}^2}{4\pi}\frac{e^{-m_\omega^* R}}{R}-A_2\frac{g_{\sigma
H}^2}{4\pi}\frac{e^{-m_\sigma^* R}}{R}\right)
\end{equation}
where $A_1=6.2$ and $A_2=8.4$ are the coefficients from accounting
all of the clusters' contributions~\citep{huang1988}.
The number density of $H$-clusters $n$ is $n=R^{-3}$, so
$\epsilon_I$ can be written as the function of $n$,
\begin{equation}
\epsilon_{I}=\frac{1}{2}n^{4/3}\left(A_1\frac{g_{\omega
H}^2}{4\pi}e^{-m^*_\omega n^{-1/3}}-A_2\frac{g_{\sigma
H}^2}{4\pi}e^{-m^*_\sigma n^{-1/3}}\right). \label{ei}
\end{equation}
Considering the Brown-Rho scaling law for $H$-particles in Eq.(\ref{BRh}),
the rest-mass energy density also depends on the number density $n$
of $H$-clusters, and the total energy density is then
\begin{equation}
\epsilon=\epsilon_I+n\cdot m_H^*, \label{e}
\end{equation}
and the pressure is
\begin{equation}
P=n^2\frac{d}{dn}\left(\frac{\epsilon}{n}\right). \label{p}
\end{equation}
At the surface of a star the pressure should be vanishing,
$P(n=n_s)=0$.
Taking $g_{\omega H}/g_{\omega N}=2$~\footnote{The relation between
$g_{\omega H}$ and $g_{\omega N}$ is in fact unknown, but the ratio of
the two quantities $g_{\omega H}/g_{\omega N}$ seems crucial in our calculations.
If the ratio is not large
enough, e.g. if we choose the ratio to be 4/3~\citep[as used in][]{Glendenning:1998wp},
$g_{\sigma H}$ would become negative in some cases (e.g. $\alpha_{BR}=0.15$ and
$n_s\leqslant 2.2\ n_0$).
This is because the rest-mass energy density, also depending on the number density,
leads to a corresponding negative pressure. How to choose the value of $g_{\omega H}/g_{\omega N}$
is still in controversy, and here we only choose one possible
value to do our calculations.}~\citep{Faessler1997},
$g_{\sigma H}$ could be derived if
we know the surface number density of $H$-clusters $n_s$ (or the
corresponding surface mass-density $\rho_s$).

\begin{figure}
\includegraphics[width=3 in]{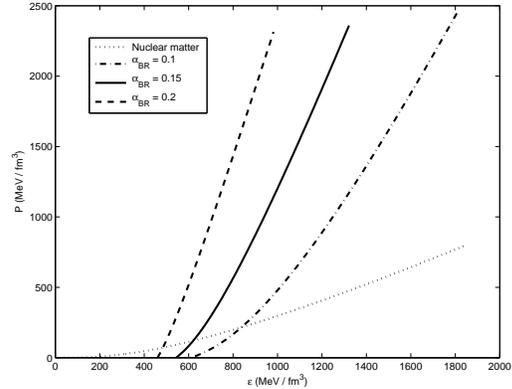}
\caption{The equation of state for nuclear
matter~\citep{NG2010} (dotted line), and $H$-cluster matter with
$\alpha_{BR}=0.1$ (dash-dotted line),
$\alpha_{BR}=0.15$ (solid line) and $\alpha_{BR}=0.2$ (dashed line).
The density ranges are the same as in Figure~\ref{figMu}.
\label{figEoS}}
\end{figure}

The equations of state of $H$-cluster matter for different cases
($\alpha_{BR}=0.1$, 0.15 and 0.2) are shown in Figure~\ref{figEoS}.
To make comparison, we also show the equation of state of nuclear
matter~\citep{NG2010}.
It is clear that the equation of state of $H$-cluster matter is much stiffer
than that of nuclear matter.
The stiffer equation of state leads to higher maximum mass, as will shown
in next subsection.
The sound speed of $H$-cluster matter with such stiff equation of state
will be discussed in \S 4.1.

%%%%%%%%%%%%%%%%%%%%%%%%%%%%%%%%%%%%%%%%
%%%%%%%%%%%%%%%%%%%%%%%%%%%%%%%%%%%%%%%%

Compact stars composed of pure $H$-clusters are electric neutral,
but in reality there could be some flavor symmetry breaking that
leads to the non-equality among $u$, $d$ and $s$, usually with less
$s$ than $u$ and $d$.
The positively charged quark matter is necessary because it allows
the existence of electrons that is crucial for us to understand the
radiative properties of pulsars.
The pressure of degenerate electrons is negligible compared to the
pressure of $H$-clusters, so the contribution of electrons to the
equation of state is negligible.

\subsection{Mass-radius relation of $H$-cluster stars}

\begin{figure}
\includegraphics[width=3 in]{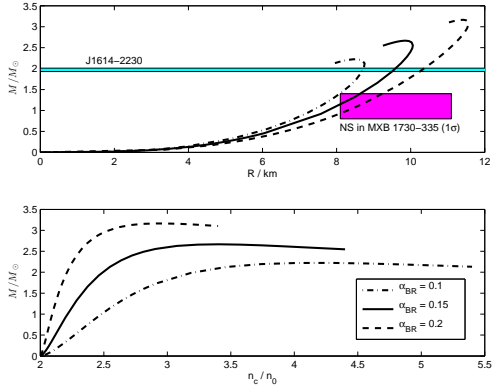}
\caption{(Color online)The mass-radius curves and mass-central
density (rest-mass energy density) curves, in the case $n_s=2
n_0$, including $\alpha_{BR}=0.1$ (dash-dotted line),
$\alpha_{BR}=0.15$ (solid line) and $\alpha_{BR}=0.2$ (dashed line).
The blue line (region) shows the
mass of PSR J1614-2230~\citep{Demorest:2010bx}, and the red
rectangle shows the mass and radius (1 $\sigma$) of the neutron
star in MXB 1730-335~\citep{sala2012}.
\label{figMR}}
\end{figure}

In general relativity, the hydrostatic equilibrium condition in
spherically symmetry is~\citep{OV1939}
\begin{equation}
\frac{1-2Gm(r)/c^2r}{P+\rho c^2}r^2 \frac{d
P}{dr}+\frac{Gm(r)}{c^2}+\frac{4\pi G}{c^4}r^3P=0,
\end{equation}
where
\begin{equation}
m(r)=\int^r_0 \rho \cdot 4\pi r^{\prime 2} dr^{\prime},
\end{equation}
with $\rho=\epsilon_{I}/c^2+m_H^* n$.
$m_H^*$ is the in-medium value for the mass of $H$-clusters, which
obeys the same scaling law as in Eq.(\ref{BRh}).
Inserting the equation of state $P(\rho)$ we can get the total mass
$M$ and radius $R$ of an $H$-cluster star by numerical integration.
Figure~\ref{figMR} shows the mass-radius and mass-central density
(rest-mass energy density) curves, in the case $n_s=2 n_0$,
including $\alpha_{BR}=0.1$, 0.15 and 0.2.
At first, $M$ grows larger as central density increases, and
eventually $M$ reaches the maximum value, after which the increase
of central density leads to gravitational instability.
In the figure, all the curves have maximum masses higher than $2
M_\odot$
%

%%%%%%%%%%%%%%%%%%%%%%%%%%%%%%%%%%%%%%%%
%%%%%%%%%%%%%%%%%%%%%%%%%%%%%%%%%%%%%%%%

%
The observed masses of pulsars put constraints on the state of quark
matter.
Quark stars have been characterized by soft equations of state,
because in conventional quark star models (e.g. MIT bag model)
quarks are treated as relativistic and weakly interacting particles.
Radio observations of a binary millisecond pulsar PSR J1614-2230
imply that the pulsar mass is 1.97$\pm$0.04
$M_\odot$~\citep{Demorest:2010bx}, shown in Figure~\ref{figMR}
as the blue region.
Although we could still obtain high maximum masses under MIT bag
model by choosing suitable parameters~\citep{zdunik2000}, a more
realistic equation of state in the density-dependent quark mass
model~\citep[e.g.,][]{dey1998} is very difficult to reach a high
enough maximum stellar mass, which was considered as possible
negative evidence for quark stars~\citep{cottam2002}.
Nevertheless, some other models of stars with quark matter could be
consistent with the observation of the high mass pulsar, such as
color-superconducting quark matter model~\citep{Ruster2004} and
hybrid star models\citep{Alford2008,Baldo:2002ju}.
Moreover, quark-cluster stars could also have maximum mass $M_{\rm
max}>2M_\odot$ because of stiff equation of state~\citep{Lai:2008cw,
LX09b, Lai:2010wf}.
Recently, the mass and radius of the neutron star in the Rapid
Burster MXB 1730-335 has been constrained to be $M=1.1\pm
0.3M_\odot$ and $R=9.6\pm 1.5$ km (1 $\sigma$) by the analysis of
{\em Swift}/XRT time-resolved spactra of the burst~\citep{sala2012},
and this result is also shown in Figure~\ref{figMR} as the red rectangle.
Our results are consistent with both observations (at least in 2 $\sigma$).

The real state of matter at densities of compact stars is
essentially a non-perturbative QCD problem and thus hard to solve.
We make a phenomenological model that quarks could be grouped into
quark-clusters at this energy scale to propose the quark-cluster
stars, and specify quark-clusters in this paper to be $H$-clusters.
The color super-conductivity model is the most often used one for
modeling quark matter, but it is still uncertain that whether the
interaction between quarks is weak enough to make the
non-perturbative treatment to be reasonable.
The point we put in this paper is that, under the assumption of
light flavor symmetry, $H$-clusters could be the possible kind of
quark clusters, and as a specific quark-cluster stars, $H$-cluster
stars could not be ruled out by the observed high mass pulsars.

\subsection{Maximum mass of $H$-cluster stars}

\begin{figure}
 \includegraphics[width=3 in]{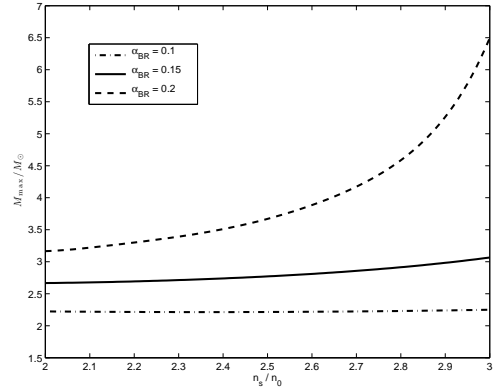}
\caption{The dependence of $M_{\rm max}$ on surface density,
in the case $\alpha_{BR}=0.1$ (dash-dotted line),
$\alpha_{BR}=0.15$ (solid line) and $\alpha_{BR}=0.2$ (dashed line).
\label{figMmax}}
\end{figure}

We constrain $\alpha_{BR}$ in the context of $H$-cluster stars by
the maximum mass of pulsars $M_{\rm max}$, shown in Figure~\ref{figMmax}.
The interaction between $H$-dibaryons was studies previously and the
related parameters were derived by fitting data in experiments of
nucleon-nucleon interaction and hypernucleus events~\citep[e.g.,
see][and references therein]{Sakai:1997jf}; however, whether the
two-particle interaction data are adequate in determining the
properties of quark matter is uncertain.
Our model for $H$-$H$ interaction could provide us another way to
study the properties of $H$-clusters in quark matter, although
giving wide ranges of parameters due to the uncertainty of $M_{\rm
max}$.
Figure~\ref{figMmax} shows the dependence of $M_{\rm max}$ on
$\alpha_{BR}$, in the case $\alpha_{BR}=0.1$, 0.15 and 0.2.
When $\alpha_{BR}\leq 0.15$, the discrepancy between different
values of surface densities is not significant, and under the range of
parameter-space we choose here $M_{\rm max}$ can be well above
$2M_\odot$.
%

%%%%%%%%%%%%%%%%%%%%%%%%%%%%%%%%%%%%%%%%
%%%%%%%%%%%%%%%%%%%%%%%%%%%%%%%%%%%%%%%%

We derive the maximum mass of $H$-cluster stars to show that they
could have safe maximum mass high enough to accord with the
observations, although there are certainly some other kind of quark
star models which provide possible ways to explain the observed high
mass of the newly discovered pulsar PSR J1614-2230. However, it
should be noted that the highest mass of pulsars that we find is
surely different from the real (theoretical) maximum mass that a
stable pulsar is able to have against gravity.
But how to inter the real maximum mass from the observed masses?
The relation between the two in the case of pulsars could be
compared to the case of white dwarfs, if we assume that
the observed mass of pulsars and white dwarfs
(or in fact any other objects) could
have the same bias towards their real maximum mass due to observational
effects.
Here we chose white dwarfs to make the comparison since
the maximum mass of white dwarfs is
well established to be about $1.4 M_\odot$.

The statistical study of
nearby white dwarfs lying within 20 pc of the Sun shows that the
distribution of measured masses of such sample of white dwarfs has a
peak at around $0.6 M_\odot$ and the most massive one is about $1.3
M_\odot$~\citep{Kepler07,Giammichlel12}. Assuming the same scaling
of measured masses and the real maximum mass for the
case of pulsars, whose distribution of measured masses shows a peak
at around $1.4 M_\odot$~\citep{zcm11,horvath11}, we could infer that
the maximum mass of pulsars can be estimated to be $\sim 3.3
M_\odot$ (using the peak value) or $\sim 2.2 M_\odot$ (using
$2M_\odot$ as the maximum value).
The estimated maximum mass for pulsar-like stars would be $\sim
3 M_\odot$, which is still much lower than the detected minimal
mass ($\sim 5M_\odot$) of stellar black holes~\citep{bh-mass}, if
the mass ($2.74M_\odot$) of a pulsar (J1748-2021B) in a globular
cluster is confirmed in the future.
As shown by our results, $H$-cluster stars are consistent with the
above estimation because their maximum mass could be $\sim 3
M_\odot$ or even higher
(e.g. $\alpha_{BR}\geq 0.15$). Therefore, discovering more massive
pulsars in the future will certainly be helpful for us to get closer
to the maximum mass and distinguish different models.

\section{Discussions}

\subsection{About the stiff equation of state}

\begin{figure}
 \includegraphics[width=3 in]{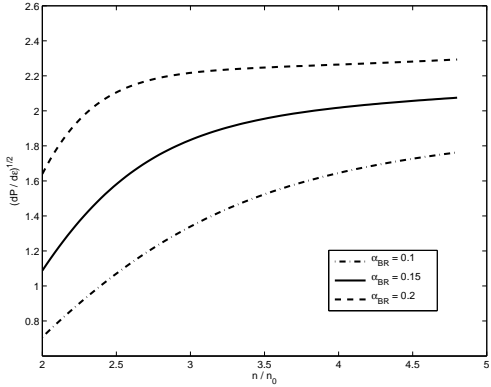}
\caption{The values of $\sqrt{dP/d\epsilon}$ inside
stars, in the cases $\alpha_{BR}=0.1$ (dash-dotted line),
$\alpha_{BR}=0.15$ (solid line) and $\alpha_{BR}=0.2$ (dashed line).
\label{figCs}}
\end{figure}

Composed of non-relativistic $H$-clusters with interaction in the
form of Eq.(\ref{V}), quark-cluster stars could have a stiff
equation of state and a high maximum mass.
Under a wide range of densities, showing in Figure~\ref{figCs},
we have $\sqrt{dP/d\epsilon}>1$.
If we treat the sound speed (i.e. the signal propagation speed)
as $c_s=\sqrt{dP/d\epsilon}$, then we
will have the ``superluminal'' sound speed.
However, whether the real sound speed can always been calculated
as $c_s$ may depend on the structure of matter and the way we
introduce the inter-particle interaction.
The probability that $c_s$ exceeding the speed of light
in ultradense matter was studied previously by~\cite{Bludman:1968zz}.
\cite{Caporaso:1979uk}
then proposed explicitly that, in a
lattice with $P(\epsilon)$ relation arising from a static calculation,
$c_s$ is not a dynamically meaningful speed, i.e., $c_s$ is not
a signal propagation speed.

%%%%%%%%%%%%%%%%%%%%%%%%%%%%%%%%%%%%%%%%
%%%%%%%%%%%%%%%%%%%%%%%%%%%%%%%%%%%%%%%%

Causality is a very basic property of dense matter theory, widely accepted.
In the case of fluid the real sound speed is calculated as $c_s$, so causality
is violated when $dP/d\epsilon >1$.
However, the situation is much more complicated when translational symmetry
breaks for solid quark-cluster matter where quarks are clustered in lattice.
Moreover, we use a classical potential model, i.e. a kind of action at a distance,
to derive the total energy density and thus the equation of state.
In this non-relativistic model, $P(\epsilon)$ may only measure the stiffness of
equation of state, and the real sound speed is too difficult to calculate.
From the picture of the mechanism for interaction in our model, the interaction
is mediated by mesons, so the real speed of
interaction is obviously smaller than the speed of light, i.e., the signal
propagation speed remains subluminal.

Our equation of state is very stiff because of two main reasons.
Firstly, quark-clusters are non-relativistic particles.
For the system composed of non-relativistic particles, ignoring the interaction,
the momentum of each particle could be approximated as $p\propto n^{1/3}$
(from Heisenberg's relation), and the total energy density $\epsilon\propto n$.
The kinetic energy per particle $E_k\propto p^2$, and then the kinetic energy density
$\epsilon_k\propto np^2 \propto n^{5/3}$,
so pressure $P=n^2\partial(\epsilon_k/n)/\partial n\propto n^{5/3}\propto \epsilon^{5/3}$.
On the other hand, for extremely relativistic particles, from the same estimation,
$P\propto \epsilon$.
Therefore, the equation of state of non-relativistic system is stiffer than that of relativistic one.
Secondly, each quark-cluster is trapped inside the potential well formed by the neighboring
quark-clusters, and the in-medium stiffening effect makes the
shape of potential well to be stiffer.
From the above arguments, our model is different from the mainstream that quark stars,
composed of relativistic and weakly
interacting quarks, are characterized by soft equation of state.
Phenomenologically, a corresponding-state approach to quark-cluster
matter~\citep{Guo12} also results in a very stiff equation of state.

\subsection{The binding of $H$-clusters inside stars}

At the highest density of the stars, with $n_B\simeq8.3n_0$,
the distance between two nearby
$H$-clusters is about 1.1 fm. The size of $H$-dibaryons could be a
little larger than that of nucleons, and then at the center of the
star they could be so crowded that they touch the nearby ones, but
they should be safe against being crushed. We assume that the touch
of nearby $H$-clusters does not influence our overall picture, since
it only happens at the very center of the star and the degree of
touch is not high to cause dissociation.

In fact, the dependence of binding energy of quark-clusters on
density is still unknown. However, if the mass of $H$-dibaryons
decreases with increasing densities like baryons and mesons, this
could be equivalent to the increasing of binding energy of
$H$-dibaryons. At densities beyond $\rho_0$, the degrees of freedom
become complex due to the non-perturbative nature of QCD, which
could be responsible to the different binding behavior to the case
at densities below $\rho_0$.

Conventional strange quark matter (without quark-clustering) is
thought to be stable in bulk but unstable in the case of light strangelets
when the baryon number $A$ is as small as 6~\citep[see a review in][]{Madsen1998}.
Although quark-clusters are similar to light strangelets, they could still be
stable as they are in medium but not in vacuum.
The highly dense environment makes the energy per baryon of $H$-clusters
to be lower than that of neutron and nucleon matter, as has shown above.
As a kind of light strangelets, $H$-dibaryons are unstable in vacuum, which make
it difficult to study them experimentally; however, inside compact stars,
the in-medium effect could stabilize them.

\subsection{$H$-cluster stars are self-bound}

It is worth noting that, although composed of $H$-clusters,
$H$-cluster stars are self-bound. They are bound by the interaction
between quark-clusters (the $H$-clusters here). This is different
from but similar to the traditional MIT bag scenario. The
interaction between $H$-clusters could be strong enough to bind the
star, and on the surface, the quark-clusters are just in the
potential well of the interaction, leading to non-vanishing density
but vanishing pressure.

It is surely possible that there could be normal matter surrounding
a self-bound $H$-cluster star, but initially the surroundings would
not remain because of energetic exploding~\citep{orv05,ph,cyx}.

\subsection{Clustering quark matter}

Quark-clusters could emerge in cold dense matter because of the
strong coupling between quarks.
The quark-clustering phase has high density and the strong
interaction is still dominant, so it is different from the usual
hadron phase.
On the other hand, the quark-clustering phase is
also different from the conventional quark matter phase which is
composed of relativistic and weakly interacting quarks.
The quark-clustering phase could be considered as an intermediate
state between hadron phase and free-quark phase, with deconfined
quarks grouped into quark-clusters, and that is another reason why
we take quark-cluster stars as a special kind of quark stars.
$H$-cluster stars are self-bound due to the interaction between
clusters, with non-vanishing surface density but vanishing surface
pressure.

Whether the chiral symmetry broken and
confinement phase transition happen simultaneously inside compact
stars is a matter of debate~\citep[see][and references
therein]{Andronic:2009gj}, but here we assume that the chiral
symmetry is broken in quark-clustering phase.

\subsection{From the asymmetry term to a flavor symmetry}

It is well know that there is an asymmetry term to account for the
observed tendency to have equal numbers of protons ($Z$) and
neutrons ($N$) in the liquid drop model of the nucleus. This nuclear
symmetry energy (or the isospin one) represents a symmetry between
proton and neutron in the nucleon degree of freedom, and is actually
that of up and down quarks in the quark degree~\citep{Li_Chen2008}.
The possibility of electrons inside a nucleus is negligible because
its radius is much smaller than the Compton wavelength $\lambda_c =
h/m_ec = 0.24\AA$. The lepton degree of freedom would then be not
significant for nucleus, but what if the nuclear radius becomes
larger and larger (even $\gg \lambda_c$)?

Electrons are inside a large or gigantic nucleus, which is the case
of compact stars.
Now there is a competition: isospin symmetry favors $Z=N$ while
lepton chemical equilibrium tends to have $Z\ll N$. The nuclear
symmetry energy $\sim 30 (Z-N)^2/A$ MeV (at saturated nuclear matter density
$\rho_0$), where $A=Z+N$, could be around 30 MeV per baryon if $N\gg
Z$. Interesting, the kinematic energy of an electron is $\sim 100$
MeV if the isospin symmetry holds in nuclear matter.
However, the situation becomes different if strangeness is included:
no electrons exist if the matter is composed by equal numbers of
light quarks of $u$, $d$, and $s$ with chemical equilibrium.
In this case, the 3-flavor symmetry, an analogy of the symmetry of
$u$ and $d$ in nucleus, may result in a ground state of matter for
gigantic nuclei. Certainly the mass different between $u$, $d$ and
$s$ quarks would also break the symmetry, but the interaction
between quarks could lower the effect of mass differences and try to
restore the symmetry. Although it is hard for us to calculate how
strong the interaction between quarks is, the non-perturbative
nature and the energy scale of the system make it reasonable to
assume that the degree of the light flavor symmetry breaking is
small, and there is a few electrons (with energy $\sim 10$ MeV).
Heavy flavors of quarks ($c$, $t$ and $b$) could not be existed if
cold matter is at only a few nuclear densities.

The above argument could be considered as an extension of the
Bodmer-Witten's conjecture. Possibly it doesn't matter whether three
flavors of quarks are free or bound.
We may thus re-define {\em strange matter} as cold dense matter with
light flavor symmetry of three flavors of $u$, $d$, and $s$ quarks.

\section{Conclusions}

We propose in this paper that the strong interaction between quarks
inside compact stars renders quarks grouped into a special kind of
quark-clusters, $H$-clusters, leading the formation of $H$-cluster
stars.
Although there are many uncertainties about the stability of
$H$-cluster matter, it could be possible that at high densities
$H$-cluster matter is stable against transforming to nucleon matter.

The equation of state of $H$-cluster stars is derived by assuming
the Yukawa form of $H$-$H$ interaction under meson-exchanges, and
the in-medium effect from Brown-Rho scaling law of meson-masses is
also taken into account.
$H$-cluster stars could have stiff equation of state, and under a
wide range of parameter-space, the maximum mass of $H$-cluster stars
can be well above 2$M_\odot$.
Furthermore, if we know about the properties of pulsars from
observations, we can get information on $H$-$H$ interaction; for
example, if a pulsar with mass larger than $3M_\odot$ is discovered,
then we can constrain the coefficient of Brown-Rho scaling
$\alpha_{BR}\geq 0.15$.

Although the state of cold quark matter at a few nuclear densities
is still an unsolved problem in low energy QCD, various pulsar
phenomena would give us some hints about the properties of elemental
strong interaction~\citep{Xu:2010}, complementary to the terrestrial
experiments.
Pulsar-like compact stars provide high density and relatively low
temperature conditions where quark matter with $H$-clusters could
exist.
$H$-cluster has been the subject of many theoretical and
experimental studies.
It is in controversy that whether $H$-cluster is a bound state,
which depends on the quark-masses~\citep{Shanahan:2011su}, and the
binding behavior at high densities is still unknown.
Whether quark matter composed of $H$-clusters could achieve at
supra-nuclear density is still uncertain, and on the other hand, the
nature of pulsar-like stars also depends on the physics of dense
matter.
These problems are essentially related to the non-perturbative QCD,
and we hope that future astrophysical observations would test the
existence of $H$-cluster stars.

Finally, we would clarify two questions and answers, which should be
beneficial to make sense about the conclusions presented in this
paper.
(1) {\em Why does not an $H$-particle on the surface decay into
nucleons?} The reason could be similar to that why a neutron does not
decay into proton in a stable nucleus. \cite{Landau1938}
demonstrated that significant gravitational energy would be released
if neutrons are concentrated in the core of a star, it is now
recognized, however, that fundamental color interaction is more
effective and stronger than gravity to confine nucleons. The equality
condition of chemical potentials at the boundary between two phases
applies for gravity-confined stars~\citep{Landau1938}, but may not
for self-bound objects by strong interaction.
(2) {\em Why can hardly normal matter be converted into more stable
$H$-cluster matter in reality?} We know $^{56}$Fe is most stable
nucleus, but it needs substantial thermal kinematic energy to make
nuclear fusion of light nuclei in order to penetrate the Coulomb
barrier. Strong gravity of an evolved massive star dominates the
electromagnetic force, compressing baryonic matter into
quark-cluster matter in astrophysics. This is expensive and rare.

{\bf Acknowledgement:}
We would like to thank useful discussions at our pulsar group of PKU. This work is supported by the National Basic Research Programme of China
(Grant Nos. 2012CB821800, 2009CB824800), the National Natural Science Foundation of China (Grant Nos. 11203018, 11225314, 10935001).

\end{document}